# Evaluation of Contemporary Convolutional Neural Network Architectures for Detecting COVID-19 from Chest Radiographs

Nikita Albert
*College of Engineering and Computer Science*
*Syracuse University*
Syracuse, NY, USA
nikitaalbert@gmail.com

*Abstract*— **Interpreting chest radiograph, a.ka. chest x-ray, images is a necessary and crucial diagnostic tool used by medical professionals to detect and identify many diseases that may plague a patient. Although the images themselves contain a wealth of valuable information, their usefulness may be limited by how well they are interpreted, especially when the reviewing radiologist may be fatigued or when or an experienced radiologist is unavailable. Research in the use of deep learning models to analyze chest radiographs yielded impressive results where, in some instances, the models outperformed practicing radiologists. Amidst the COVID-19 pandemic, researchers have explored and proposed the use of said deep models to detect COVID-19 infections from radiographs as a possible way to help ease the strain on medical resources. In this study, we train and evaluate three model architectures, proposed for chest radiograph analysis, under varying conditions, find issues that discount the impressive model performances proposed by contemporary studies on this subject, and propose methodologies to train models that yield more reliable results.. Code, scripts, pre-trained models, and visualizations are available at** https://github.com/n-albert/COVID-detection-from-radiographs.

## I. INTRODUCTION

The COVID-19 pandemic has disrupted the world in an unimaginable way with daily routines put on hold and every system pushed to its limits. Medical institutions have faced the greatest strain, with the sheer number of COVID-19 patients inundating nurses, doctors, and hospitals around the world. The systemic deficits, such as limited resources, underfunding, etc., have made the situation so dire that overwhelmed hospitals float the idea of resorting to triage [1] and raise the fear that this pandemic may not be effectively contained.

Advancements in deep learning have inspired researchers to explore how said advancements may be used to develop algorithms, models, and systems for affordable, effective medical diagnostic tools that can help alleviate some of the aforementioned deficits. Researchers at Stanford University, for example, proposed CheXNet [2], a 121-layer convolutional neural network, that was meant to detect pathologies from chest radiographs accurately and avoid errors that an overworked or undertrained radiologist may make. The model was trained on the ChestX-ray14 dataset [3] and was able to detect 14 different pathologies at rates exceeding those of practicing radiologists.

Although such model performance is indeed impressive and promising, the CheXNet model, and models similar to it, carry inherent disadvantages. Specifically, their significant depth and size require significant amounts of memory and computational resources for training and prediction. They also possess a large number of trainable parameters, in the order of hundreds of thousands, which make the models prone to overfitting and poor at generalization, especially when applied on datasets with limited observations.

These issues, however, do not prove entirely insurmountable and instead motivate research in developing more efficient, "lightweight" models that make use of critical features from their deep counterparts. Researchers at the Technical University of Munich, for example, sought to address said resource issues by proposing a neural network, composed of 5 convolutional blocks utilizing shortcut connections, a global average pooling layer and a fully connected softmax layer [4]. They trained this model to detect whether a radiograph is affected by tuberculosis or not on the Montgomery and Shenzhen datasets [5] and achieved predictive results comparable to other publications while using significantly fewer computational, memory and power resources

Although the aforementioned models may differ structurally, they attempt to identify and learn from the unique features manifested by different pathologies on a radiograph [6] [7] to perform the same diagnostic analyses done by radiologists. With the magnitude of this crisis, researchers and professionals have worked to compile as much data as possible, including radiographs, to better understand the disease; For example, it is now evident that COVID-19 manifests unique radiographic features [8] [9], such as the notorious "ground-glass" opacity. Some research, in the use of deep models to detect COVID-19, has been done and offers very impressive results; In some cases, the proposed models have detected COVID-19 infections perfectly, with no misclassifications.

In this paper, we aim to explore how existing model architectures, that performed well at their respective chest radiograph diagnostic tasks, fare at learning and detecting COVID-19 infections. We will, specifically, explore how each architecture handles dealing with a dataset with limited observations, how we may use data augmentation to improve model performance, and evaluate whether current methodologies for this task are sufficient enough to yield robust results.

## II. ARCHITECTURES OVERVIEW

In this section, we detail the key features, motivations, and past performances of the model architectures we evaluate in this study.

## A. DenseNet-121

The *Dense Convolutional Network,* or *DenseNet*, architecture was proposed in 2016 Huang et al. [10], researchers from Cornell University, Tsinghua University, and the Facebook AI Research Group. The main motivation was to develop an architecture that addresses the "vanishing gradient" problem and allows for significantly deeper models.

The "vanishing gradient" problem typically an input passes through many layers of a deep neural network and may vanish or "wash-out" by the time it reaches the end. This leads to the resultant gradient to become too small have any affect on the weights and biases of initial layers during training, and subsequently causes overall model training to fail.

*DenseNet* addresses this problem by connecting all layers (with matching feature-map sizes) directly with each other so that each layer obtains additional inputs from all preceding layers and passes on its own feature-maps to all subsequent layers to ensure maximum information flow between layers in the network. This allows *DenseNet* to explicitly differentiate between what information is added to the network and what information is preserved between layers. This layout is illustrated below, in *Figure 1*.

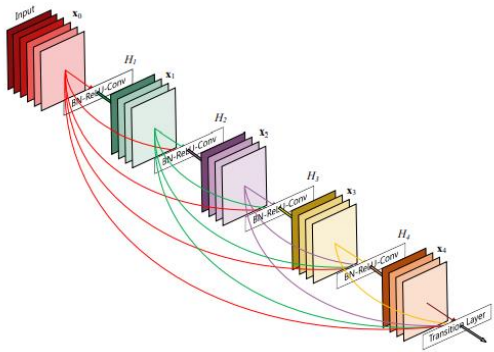

*Figure 1. Example of 5-layer DenseNet model.*

The proposed architecture was evaluated on the CIFAR-10, CIFAR-100, SVHN, and ImageNet object recognition benchmark tests and obtained significant improvements over the prior state-of-the-art results on most of them. This model architecture, pretrained on the ImageNet benchmark, was utilized by Rajpurkar et al. to develop the CheXNet algorithm [2], which was able to achieve state-of-the-art results at detecting 14 different pathologies from chest radiographs and at rates exceeding those of practicing radiologists.

We choose to include the *DenseNet-121* architecture in our study because of such promising performance on benchmark and similar diagnostic tasks, and to evaluate how it may fare at detecting COVID-19 infections.

## B. ResNet-50

Prior to the proposal of the *DenseNet* architecture, researchers at Microsoft, He et al., proposed the *deep residual learning* framework [11], or *ResNet*, which first introduced the idea of using shortcut connections to solve the vanishing gradient problem. The guiding design principle of this framework was to have layers within the network fit a provided residual mapping, instead of hoping that every few layers would fit a desired underlying mapping.

This was realized through the use of "shortcut connections" to perform identity mapping of an input, adding the mapping to the output of some stacked layers and having that summation be the input to subsequent layers. The hypothesis was it would be easier for such model to train and optimize towards this residual mapping than to optimize towards an unreferenced mapping.

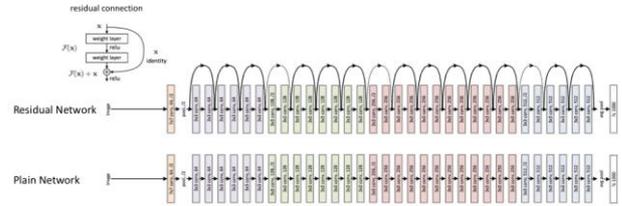

*Figure 2. Comparison of Residual Network vs. Normal Network*

This framework was the foundation of the team's submissions to the ILSVRC and COCO 2015 competitions, which achieved 1st place on the tasks of ImageNet detection, ImageNet localization, COCO detection, and COCO segmentation.

Since then, *ResNet* models have been applied extensively towards image classification, segmentation, and localization tasks, especially in the medical domain. In Wang et al., the *ResNet-50* model scored performed closely to their proposed architecture, with a positive predictive value of 98.8% for detecting COVID-19 [12]. Farooq et al. [13] even managed to achieve a positive predictive value of 100% for detecting COVID-19.

We choose to include the *ResNet-50* model architecture in this study due to such past performance and promise.

## C. "Efficient" Model

So far, the models we have selected have been very deep convolutional neural networks that were originally developed for natural image classification on datasets compromising up to millions of images and thousands of target labels. Although these models have yielded great results when applied in the medical domain, they carry intrinsic disadvantages that may make them untenable to use under certain conditions. These models, for example, have a large number of parameters, high hardware requirements and, therefore, could be prone to overfitting or deploy in mobile settings.

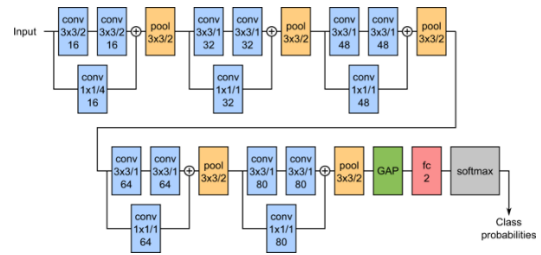

*Figure 3. Schematic representation of the Efficient network architecture*

Researchers at the Technical University of Munich, Pasa et al. [4], proposed a simple model architecture that is faster and more efficient than deep models like our aforementioned ones, while achieving comparably impressive accuracy detecting tuberculosis from chest radiographs. The architecture is similar to that of the *ResNet*, with the key difference of having the shortcut connections perform 1x1 convolution prior to summation and max pooling is performed before each convolutional block. When applied on the combined Montgomery and Shenzhen dataset [5], the proposed model achieved a 92.5% accuracy at detecting whether an individual is suffering from tuberculosis or not, while achieving significantly quicker training and inference times than other contemporary proposals.

We choose to include this model in this study to see if it possible to apply it towards detecting COVID-19 infections and compare its efficacy against the other two architectures, especially when we expect it to use significantly fewer resources.

### III. DATASET

The dataset we use is the *CoronaHack – Chest X-Ray Dataset* from Kaggle [14] accessed on April 22, 2020. This dataset is a collection of chest radiographs of healthy and pneumonia afflicted patients, with samples attributing the pneumonia to viral, bacterial, coronavirus, etc. causes. It is essentially a pared down version of University of Montreal's Postdoctoral Fellow Joseph Paul Cohen's *COVID-ChestXRay-Dataset* on Github [15], which is currently widely accessed and utilized for contemporary studies. As such, we feel confident in the use of this dataset in the study.

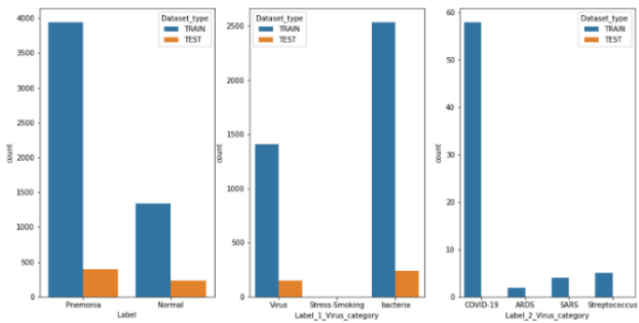

*Figure 4. Initial Class Distribution of the Dataset*

The only issues we have is that the dataset is very unbalanced, as seen in *Figure 4*, and some classes do not have any samples allocated for test yet. To bring the dataset into a more workable form, we make some design choices and manipulate the dataset accordingly. First, we drop any samples labelled *Stress-Smoking*, *ARDS* or *Streptococcus,* as those labels contain less than ten observations each. Although we also have few observations of the *SARS* label, we choose to consolidate those samples with the *COVID-19* samples under the new label *COVID*, as both diseases are due to the SARS-CoV-*X* pathogens and manifest similar symptoms [16].

We then sample 1400 observations labelled *Normal*, 700 observations labelled *Pneumonia* and *Virus,* 700 observations labelled *Pneumonia* and *Bacteria*, and all observations labelled *COVID*. We subsequently split each of these sample groups by a ratio of 0.64-0.16-0.20 and rejoin the resulting splits to form our training, validation, and testing sets, respectively; This ensures we have samples for each target label in each of our sets.

Finally, as we generate a copy of our rebalanced dataset, we resize all of our images to be of dimensions 768x768 pixels. We display our new working dataset distribution below, in *Figure 5*.

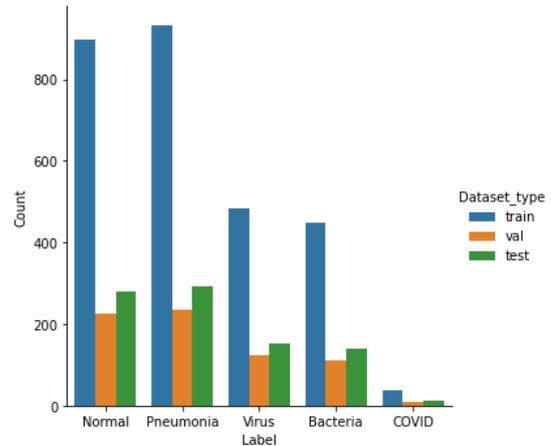

*Figure 5. Rebalanced Dataset Class Distribution*

### IV. EXPERIMENTS OVERVIEW

In the studies proposing and utilizing the aforementioned architectures, the models were trained to perform discrete multi-class classification:

- CheXNet (*DenseNet-121*): An input radiograph can only be classified as being affected by 1 out of 14 diseases [2].
- *ResNet-50*: An input can only be classified as normal, affected by pneumonia or affected by COVID in Wang et al.'s study [12].
- *Efficient* Model: An input is either affected by tuberculosis or not.

Although this design choice makes sense for certain goals and datasets, we believe that it should not be made or implemented on models meant to detect the presence of COVID-19; COVID-19 was originally described as a disease that was no worse than the flu, but has confounded the medical community with the emergence of Kawasaki's disease in children, strokes in young adults and severe respiratory function decline in others. Simply, we still do not know and understand how COVID-19 manifests itself in and affects individuals. There is also mounting evidence that the COVID-19 coinfection rate is close to 21%, which makes it so that "identification of another pathogen may not rule out the presence of the novel coronavirus" [17].

To ensure our models can be scaled up to be trained on more expansive datasets than the one used in this study and do not rule out COVID-19 infection due to the detection of another infection, and vice-versa, we make the design choice of having our final fully-connected layers use 'sigmoid' activation, to perform multi-label classification, instead of 'softmax' activation (multi-class classification).

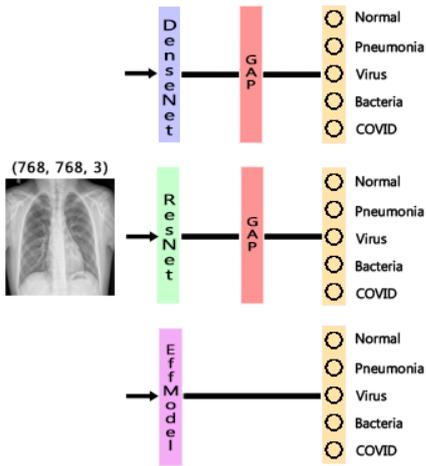

*Figure 6. High Level Overview of Model Structures*

We also include a global averaging pool layer before the fully connected layer in our *DenseNet* and *ResNet* models. Zhou et al. have shown in their study [18] that the inclusion of this layer can yield valuable information on the object localization done by a neural networks, even ones trained for just classification; We include this layer to assist in subsequent visualization.

To fully evaluate the three aforementioned architectures, we perform three rounds of training, inference and visualization under different model conditions and inputs. The methodologies and justifications are detailed below.

### A. Experiment A

In contemporary studies, exploring the use of machine learning models in detecting COVID-19 from chest radiographs, (Ilyas et al. [19], Mangal et al. [20], Wang et al. [12], Farooq et al. [13]) there are no mentions of performing any type of preprocessing on dataset or individual images, aside from correcting class imbalances. In *Experiment A*, we aim to follow this methodology by training and testing our models on our aforementioned sampled dataset; We perform no modifications or processing on the images, aside from scaling the pixel values to be from 0 to 1 prior to feeding them to the our models.

We also train two versions of *DenseNet-121* and *ResNet-50* models: one using pretrained weights from the ImageNet dataset and one without any loaded weights. We do this to evaluate how the inclusion/exclusion of pretrained weights affect training, predictive and feature localization performance and also provide a fair shake for the *Efficient* model, which was proposed without being pretrained on any dataset.

We train all models in this experiment for 30 epochs with the Adam+AMSGrad optimizer with the following parameters: $\beta_1 = 0.9$, $\beta_2 = 0.999$, and learning rate = $1\times10^{-4}$.

During training we also make use of callback functions to reduce our models' learning rates on plateaus and stop training early if there is no improvement in validation loss after a certain number of epochs. We set the learning rate callback to reduce the learning rate by a factor of 0.5, if no improvement is seen after 3 epochs, up to a minimum learning rate of $1\times10^{-8}$. We set the early stopping callback to stop model training if there is no improvement in validation loss after 12 epochs.

### B. Experiment B

Although we somewhat rebalanced our dataset, we still have only 36 COVID labelled samples to train on, 10 samples to validate on, and 12 samples to test on. We are concerned that this imbalance would provide insufficient information to our models and end up training them to detect COVID in an undesired way, such through identifying some commonality that is not some feature within the lung itself.

In this experiment, we attempt to address this deficit generating more samples of COVID radiographs through the use of image augmentation. We use the *ImageDataGenerator* [21] function from the *Keras* library to automatically generate new randomly augmented images from every COVID labeled image.

We set our augmentation parameters for our training and validation sets to:

- height_shift_range = 0.05
- rotation_range = 5
- horizontal_flip = True
- brightness_range = [0.9, 1.1]
- zoom_range = [0.9, 1.1]

We set our augmentation parameters for our testing set to:

- height_shift_range = 0.05
- rotation_range = 2.5
- horizontal_flip = True
- brightness_range = [0.95, 1.05]
- zoom_range = [0.95, 1.05]

For every COVID image in our training and validation sets, we generate 15 augmented images; For every COVID image in our testing set, we generate 12 augmented images. We generate a new dataset with these new images and the base images for the other target labels for accessibility and subsequent use in visualization. We ensure that new augmented images remain in their respective and intended datasets, so that an augmented image of a training image does not end up in the augmented testing set and such.

In this experiment, unlike in *Experiment A*, we train just one version of the *DenseNet-121* and *ResNet-50* models. Specifically, the versions with from *Experiment A* that perform the best, i.e. the versions with or without weights from ImageNet.

We train all models in this experiment for 30 epochs with the Adam+AMSGrad optimizer with the following parameters: $\beta_1 = 0.9$, $\beta_2 = 0.999$, and learning rate = $1\times10^{-4}$. We also maintain the use of the same callback functions and callback settings from *Experiment A*.

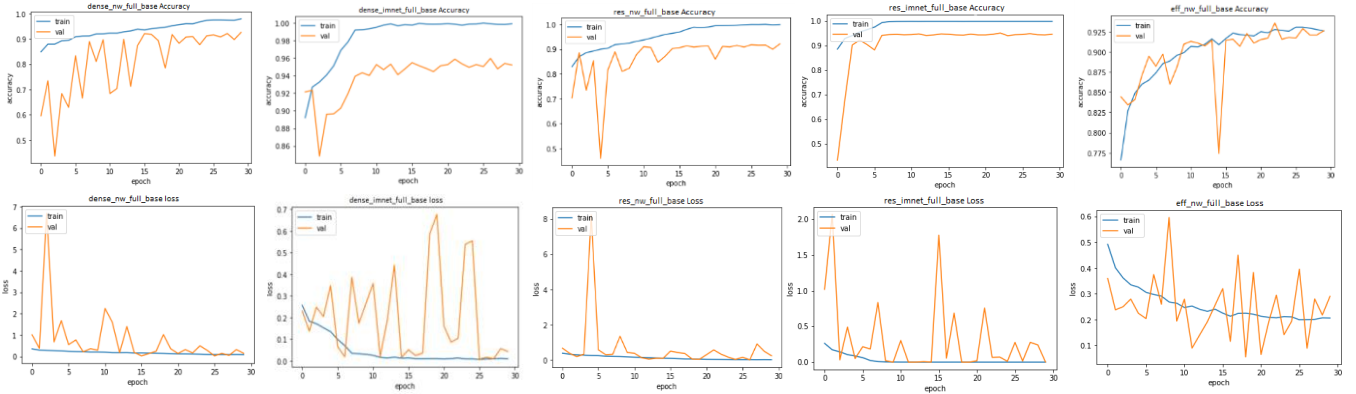

*Figure 8. Top Row: Model training and validation accuracy history*
*Bottom Row: Model training and validation loss history*
*Left to Right: Unweighted DenseNet, ImageNet DenseNet, Unweighted ResNet, ImageNet ResNet, Efficient model*

## C. Experiment C

In such studies, significant discussion is typically afforded to the hypothesized or necessary preprocessing done to a dataset in order to properly train well-performing, generalized models. We find it strange that little mention is given to such preprocessing in contemporary studies on applying machine learning to detect COVID infection from radiographs.

To address this deficit, we perform preprocessing similar to the preprocessing done by Pasa et al. to train their proposed *Efficient* model [4]. First, we crop the 40 pixels off the top and left sides and 30 pixels off the bottom and right sides of each image. Then, we resize each image back to original size of 768 x 768 pixels. Finally, and the key difference, is that we use Contrast Limited Adaptive Histogram Equalization to 'enhance' each image. We elect to use this method, instead of subtracting the mean of all pixels in the dataset from an image and dividing them by their standard deviation, to avoid relying on a methodology that requires loading the entirety of a dataset into memory, especially when a future dataset may be too large to make such methodology possible.

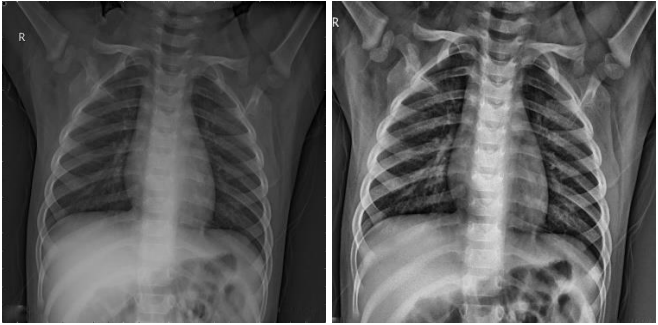

*Figure 7. Example of CLAHE adjustment on a chest radiograph*

We also apply image augmentation on all images fed into our models during training. We set our augmentation parameters to:

- rotation_range = 10
- height_shift_range = 0.05
- width_shift_range = 0.05
- shear_range = 0.02
- horizontal_flip = True

To ensure our models make the most out of augmentation, we multiply the number of steps per epoch by 2. We then train all models in this experiment for 50 epochs with the Adam+AMSGrad optimizer with the following parameters: $\beta_1 = 0.9$, $\beta_2 = 0.999$, and learning rate = $1 \times 10^{-4}$.

We continue to make use of the same callback functions; Except we modify them so that the learning rate callback reduces after 5 epochs and the early stopping callback stops training after 15 epochs without improvement in validation loss.

## V. TRAINING

We perform all model training on an Amazon Web Services Sagemaker kernel running on a *ml.g4dn.xlarge* EC2 instance. This specific EC2 instance runs with access to 1 NVIDIA T4 Tensor Core GPU, 4 Intel Xeon Scalable vCPUs and 16 GiB of RAM. Below, we quantify certain training metrics, such as training time, loss, etc., and share our initial impressions.

### A. Experiment A

We visualize our model training accuracies and losses in *Figure 8,* and we find some hints of how the models will perform post-training. For example, we find that the unweighted *DenseNet* and *ResNet* models experience higher spikes in validation loss during training than their ImageNet weights loaded counterparts. The ImageNet weighted models, however, experience continued spikes in validation loss throughout training and do not seem to fully converge towards an optimum; Despite these spikes, the weighted models achieve consistently higher validation accuracies and hint they will outperform the unweighted counterparts. We also have a strong inclination to assume that the *Efficient* model will be the worst-performing model, as both its validation accuracy and loss performances do not fully flatten out and are empirically worse than the other models'.

| Model | Time Per Epoch | Total Training Time |
|---|---|---|
| DenseNet-121 Weights/None | ~ 315 s | ~ 9450 s (2h 37m 30s) |
| ResNet-50 Weights/None | ~ 300 s | ~ 9000 s (2h 30m 0s) |
| Efficient Model | ~ 50 s | ~ 1500 s (0h 25m 0s) |

*Table 1. Experiment A model training times*

We find that the *DenseNet* and *ResNet* models achieve similar total and per epoch training times, with a 5% difference between the two. This is rather impressive for the *DenseNet* model is comprised of 121 layers and achieves a comparable training time to the *ResNet* model, which is comprised of just 50 layers. The *Efficient* model, however, achieves significantly faster training times that are 1/6th of the deeper models. Although, we hypothesize that the *Efficient* model will not perform as well as the other two, the significantly quicker training times and associated smaller resource usage may steer a design choice to select said or similar model.

### B. Experiment B

In *Experiment B*, we end up using the ImageNet weighted versions of the *DenseNet* and *ResNet* models, over the unweighted versions, as they achieve the best performance; We share said results from *Experiment A* later on.

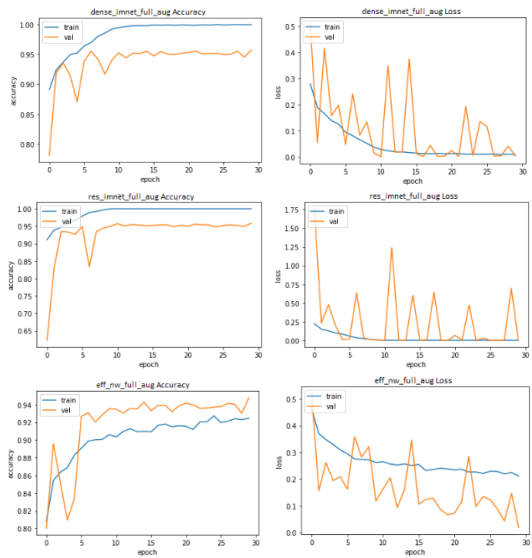

*Figure 9. Left Column: Model training and validation accuracy history*
*Right Column: Model training and validation loss history*
*Top to Bottom: ImageNet DenseNet, ImageNet ResNet, Efficient model*

Although the weighted versions of said models struggled with convergence in *Experiment A*, we find that they fare in training during *Experiment B*; This is most likely due to utilizing a better-balanced dataset. We also observe how the spikes in validation loss for the *DenseNet* and *ResNet* models are significantly lower in magnitude and frequency. With the *Efficient* model, not only do we see improved behavior with validation loss experiencing fewer peaks and small magnitudes, we also find a validation loss that is higher than the training loss.

Such improvements in the model training suggest that we will most likely find improved metrics in model performance for all models this in this experiment.

| Model | Time Per Epoch | Total Training Time |
|---|---|---|
| DenseNet-121 ImageNet | ~ 350 s | 10,500 s (2h 55m 0s) |
| ResNet-50 ImageNet | ~ 344 s | 10,320 s (2h 52m 0s) |
| Efficient Model | ~ 60 s | 1,800 s (0h 30m 0s) |

*Table 2. Experiment B model training times*

We also observe that the *Efficient* model still achieves the quickest training time, with all models experiencing some increase in training time due to the expanded, rebalanced dataset.

### C. Experiment C

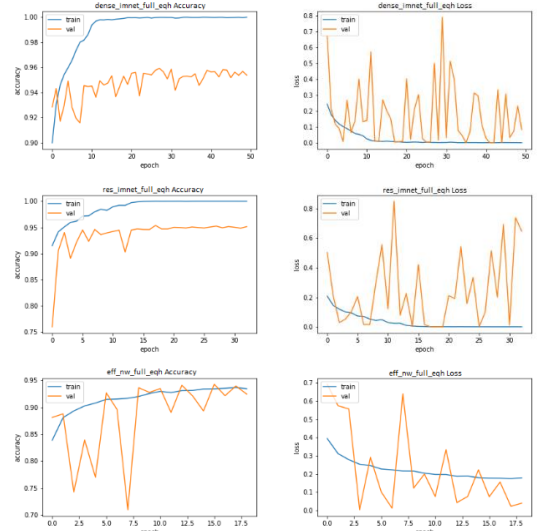

*Figure 10.. Left Column: Model training and validation accuracy history*
*Right Column: Model training and validation loss history*
*Top to Bottom: ImageNet DenseNet, ImageNet ResNet, Efficient model*

Compared to slight training improvements we saw in *Experiment B*, the use of a mildly preprocessed and heavily augmented dataset in this experiment both increased the frequency of validation loss spikes and degraded the flatness of validation accuracy for all models. Despite this, validation accuracy reaches ~94% or higher for all models at the end of training, which indicates that models perform well, however, may be struggling with handling certain targets.

| Model | Time Per Epoch | Total Training Time |
|---|---|---|
| DenseNet-121 ImageNet | ~ 1510 s | 75,500 s (20h 58m 0s) |
| ResNet-50 ImageNet | ~ 1430 s | 50,050 s (13h 54m 10s) |
| Efficient Model | ~ 550 s | 9900 s (2h 45m 0s) |

*Table 3. Experiment A model training times*

Due to the fact we use heavy augmentation on all images as we feed them to our models during training and we double the steps taken per epoch, our models see significant increases in training time; Specifically, our *DenseNet* and *ResNet* models see a ~400% increase in time per epoch, and our *Efficient* model sees a 1,000% increase in time per epoch.

We also find that only our *DenseNet* model trains for the full 50 epochs; The *ResNet* and *Efficient* models stop training after the 35th and 18th epoch, respectively, due to our early stopping callback functions kicking in. Although the validation accuracies seem to steady off around said epochs, we view such behavior with some hesitation as we may have seen better convergence after training for longer.

# VI. RESULTS

## A. Experiment A

| Model | Accuracy (95% CI) | Macro-Avg Precision (95% CI) | Macro-Avg Recall (95% CI) | Macro-Avg F1 (95% CI) | Hamming Loss (95% CI) |
|---|---|---|---|---|---|
| DenseNet-121 No Weights | 82.34% ± 3.13% | 90.34% ± 2.42% | 89.38% ± 2.52% | 89.72% ± 2.49% | 0.0622 ± 0.0198 |
| DenseNet-121 ImageNet | 84.44% ± 2.97% | 89.3% ± 2.53% | 88.93% ± 2.57% | 89.10% ± 2.55% | 0.0612 ± 0.0196 |
| ResNet-50 No Weights | 79.02% ± 3.34% | 83.25% ± 3.06% | 78.48% ± 3.37% | 80.71% ± 3.23% | 0.0832 ± 0.0226 |
| ResNet-50 ImageNet | 84.27% ± 2.99% | 88.31% ± 2.63% | 89.20% ± 2.54% | 88.71% ± 2.59% | 0.0594 ± 0.0193 |
| Efficient Model | 80.10% ± 3.27% | 69.19% ± 3.78% | 67.23% ± 3.85% | 68.18% ± 3.82% | 0.0703 ± 0.0210 |

*Table 4. Overall accuracy and macro-average metrics of Experiment A model performances*

Immediately, from the overall model metrics, we receive a clearer understanding of the hints of model behavior we observed during training. For example, we observe how the *ResNet* and *DenseNet* models with the ImageNet weights converged quicker during training and subsequently achieved higher metrics over their unweighted counterparts. This initially seems to be a confirmation of the benefits of transfer learning, where using knowledge from solving one problem, a model is able to better suited to solve a different, but related, problem.

We also observe each model achieves a relatively good 'harsh' accuracy; a metric that captures how well a model predicts all target labels of an input correctly. We also note how the macro-average metrics are higher than the 'harsh' accuracy for all models, except for the 'Efficient' model. Note: we use macro-average metrics due to the fact we have an unbalanced dataset and we want to capture how the discrete metrics of each target affect the overall metrics.

| Target | Model | Precision (95% CI) | Recall (95% CI) | F1 (95% CI) |
|---|---|---|---|---|
| Normal | DenseNet-121 No Weights | 98.52% ± 1.41% | 95.36% ± 2.46% | 96.91% ± 2.03% |
| | DenseNet-121 ImageNet | 96.48% ± 2.16% | 97.86% ± 1.70% | 97.16% ± 1.95% |
| | ResNet-50 No Weights | 93.20% ± 2.95% | 97.86% ± 1.70% | 95.47% ± 2.44% |
| | ResNet-50 ImageNet | 97.12% ± 1.96% | 96.43% ± 2.17% | 96.77% ± 2.07% |
| | Efficient Model | 96.42% ± 2.18% | 96.07% ± 2.28% | 96.24% ± 2.23% |
| Pneumonia | DenseNet-121 No Weights | 95.68% ± 2.38% | 98.63% ± 1.36% | 97.13% ± 1.96% |
| | DenseNet-121 ImageNet | 97.57% ± 1.80% | 96.23% ± 2.23% | 96.90% ± 2.03% |
| | ResNet-50 No Weights | 97.83% ± 1.71% | 92.81% ± 3.03% | 95.25% ± 2.49% |
| | ResNet-50 ImageNet | 96.61% ± 2.12% | 97.60% ± 1.79% | 97.10% ± 1.97% |
| | Efficient Model | 96.23% ± 2.23% | 96.23% ± 2.23% | 96.23% ± 2.23% |
| Virus | DenseNet-121 No Weights | 79.85% ± 6.64% | 70.39% ± 7.56% | 74.83% ± 7.19% |
| | DenseNet-121 ImageNet | 75.00% ± 7.17% | 76.97% ± 6.97% | 75.97% ± 7.08% |
| | ResNet-50 No Weights | 71.64% ± 7.47% | 63.16% ± 7.99% | 67.13% ± 7.78% |
| | ResNet-50 ImageNet | 75.97% ± 7.08% | 76.97% ± 6.97% | 76.47% ± 7.03% |
| | Efficient Model | 76.22% ± 7.05% | 71.71% ± 7.46% | 73.90% ± 7.28% |
| Bacteria | DenseNet-121 No Weights | 73.38% ± 7.32% | 80.71% ± 6.54% | 76.87% ± 6.98% |
| | DenseNet-121 ImageNet | 77.44% ± 6.92% | 73.57% ± 7.30% | 75.46% ± 7.13% |
| | ResNet-50 No Weights | 71.77% ± 7.46% | 63.57% ± 7.97% | 67.42% ± 7.76% |
| | ResNet-50 ImageNet | 79.55% ± 6.68% | 75.00% ± 7.17% | 77.21% ± 6.95% |
| | Efficient Model | 77.10% ± 6.96% | 72.14% ± 7.43% | 74.54% ± 7.22% |
| COVID | DenseNet-121 No Weights | 100% ± 0% | 66.67% ± 26.67% | 80.00% ± 22.63% |
| | DenseNet-121 ImageNet | 100% ± 0% | 100% ± 0.00% | 100% ± 0.00% |
| | ResNet-50 No Weights | 81.82% ± 21.82% | 75.00% ± 24.50% | 78.26% ± 23.34% |
| | ResNet-50 ImageNet | 92.31% ± 15.07% | 100% ± 0.00% | 96.00% ± 11.09% |
| | Efficient Model | 0.00% ± 0.00% | 0.00% ± 0.00% | 0.00% ± 0.00% |

*Table 5. Class-specific metrics of Experiment A model performances*

When we look at the class specific metrics, we find that all models achieve impressive metrics at detecting whether a chest radiograph is *Normal* or affected by *Pneumonia*. We do, however, see that all models struggle with differentiating whether a *Pneumonia* radiograph is due to *Virus* or *Bacteria*. If we analyze the recall scores, we find that our unweighted models also tend to skew to having *Pneumonia* cases be labeled as due to *Bacteria*, whereas our weighted models are able manage to better classify for both *Virus* and *Bacteria* labels.

We also find that the reason our *Efficient* model achieves a relatively good 'harsh' accuracy with poor macro-average metrics, is due to the fact that it completely fails to detect *COVID* and struggles to differentiate between *Virus* and *Bacteria* labels, as seen from those class-specific recall scores.

When we check the *Efficient* model predictions, however, we find that the model correctly predicted the *Pneumonia* and *Virus* labels for the *COVID* observations. We view this positively as the *Efficient* model just missed predicting one label for these observations; We believe that with more *COVID* observations, the model would be better able to differentiate between observations labeled *Pneumonia, Virus,* and *COVID* and observations just labeled *Pneumonia* and *Virus.*

Overall, we find that the *DenseNet-121* architecture loaded with ImageNet weights achieves the best performance and seems best suited for this task, especially with a 100% accuracy at detecting *COVID* without any misclassification. Also, as the models using loaded ImageNet weights outperform those without any pretrained weights, we choose to move forward with ImageNet loaded models for *Experiments B* and *C*.

## B. Experiment B

| Model | Accuracy (95% CI) | Macro-Avg Precision (95% CI) | Macro-Avg Recall (95% CI) | Macro-Avg F1 (95% CI) | Hamming Loss (95% CI) |
|---|---|---|---|---|---|
| DenseNet-121 ImageNet | 87.22% ± 2.47% | 91.69% ± 2.04% | 90.96% ± 2.12% | 91.30% ± 2.08% | 0.0514 ± 0.0163 |
| ResNet-50 ImageNet | 86.65% ± 2.51% | 92.51% ± 1.94% | 90.72% ± 2.14% | 91.58% ± 2.05% | 0.0491 ± 0.0160 |
| Efficient Model | 80.26% ± 2.94% | 88.90% ± 2.32% | 88.69% ± 2.34% | 88.57% ± 2.35% | 0.0719 ± 0.0190 |

Table 6. Overall accuracy and macro-average metrics of Experiment B model performances

With a more rebalanced dataset containing augmented *COVID* images, we see improvements across the board for all of our models; We see the most significant improvement with the *Efficient* model, that achieves ~88% macro-average metrics over the ~67-69% metrics we observe in *Experiment A*. From this general improvement, we can conclude that with a more mature dataset, with more "natural" *COVID* samples, we can be afforded more leeway in selecting a model that best meets other technical specifications.

| Target | Model | Precision (95% CI) | Recall (95% CI) | F1 (95% CI) |
|---|---|---|---|---|
| Normal | DenseNet-121 ImageNet | 95.14% ± 2.52% | 97.86% ± 1.70% | 96.48% ± 2.16% |
| | ResNet-50 ImageNet | 97.16% ± 1.95% | 97.86% ± 1.70% | 97.51% ± 1.83% |
| | Efficient Model | 98.41% ± 1.47% | 88.57% ± 3.73% | 93.23% ± 2.94% |
| Pneumonia | DenseNet-121 ImageNet | 98.56% ± 1.13% | 96.70% ± 1.70% | 97.62% ± 1.45% |
| | ResNet-50 ImageNet | 98.34% ± 1.22% | 97.88% ± 1.37% | 98.11% ± 1.30% |
| | Efficient Model | 92.51% ± 2.51% | 99.06% ± 0.92% | 95.67% ± 1.94% |
| Virus | DenseNet-121 ImageNet | 89.55% ± 3.56% | 84.51% ± 4.21% | 86.70% ± 3.95% |
| | ResNet-50 ImageNet | 89.43% ± 3.58% | 83.45% ± 4.32% | 86.34% ± 3.99% |
| | Efficient Model | 86.08% ± 4.03% | 82.75% ± 4.39% | 84.38% ± 4.22% |
| Bacteria | DenseNet-121 ImageNet | 75.18% ± 7.16% | 75.71% ± 7.10% | 75.44% ± 7.13% |
| | ResNet-50 ImageNet | 77.62% ± 6.90% | 79.29% ± 6.71% | 78.45% ± 6.81% |
| | Efficient Model | 68.26% ± 7.71% | 81.43% ± 6.44% | 74.27% ± 7.24% |
| COVID | DenseNet-121 ImageNet | 100% ± 0.00% | 100% ± 0.00% | 100% ± 0.00% |
| | ResNet-50 ImageNet | 100% ± 0.00% | 95.14% ± 3.51% | 97.51% ± 2.55% |
| | Efficient Model | 99.25% ± 1.41% | 91.67% ± 4.51% | 95.31% ± 3.45% |

Table 7. Class-specific metrics of Experiment B model performances

When looking at the target-specific metrics, we find significant improvements in detecting all targets, except for the *Bacteria* target, by all models; This makes sense as our models are able to train on more *COVID* samples, which are all also labeled as *Virus*. Our *Efficient* model specifically benefits the most and achieves >90% *COVID*-specific metrics, compared to failing to detect any *COVID* labels in *Experiment A*.

Although the *DenseNet* model still achieves the best performance at detecting *COVID* radiographs without any misclassifications, we believe that the *ResNet* model is the best performing model, holistically, as it achieves higher target-specific performance for all targets.

## C. Experiment C

| Model | Accuracy (95% CI) | Macro-Avg Precision (95% CI) | Macro-Avg Recall (95% CI) | Macro-Avg F1 (95% CI) | Hamming Loss (95% CI) |
|---|---|---|---|---|---|
| DenseNet-121 ImageNet | 84.38% ± 2.68% | 92.30% ± 1.97% | 88.53% ± 2.36% | 90.33% ± 2.18% | 0.0560 ± 0.0170 |
| ResNet-50 ImageNet | 83.66% ± 2.73% | 90.35% ± 2.18% | 87.89% ± 2.41% | 89.06% ± 2.31% | 0.0642 ± 0.0181 |
| Efficient Model | 82.39% ± 2.81% | 89.89% ± 2.23% | 85.76% ± 2.58% | 87.70% ± 2.43% | 0.0722 ± 0.0191 |

Table 8. Overall accuracy and macro-average metrics of Experiment C model performances

With a moderately preprocessed dataset, with all images augmented during training, we find that our models experience some degradation in performance. For example, the 'harsh' accuracies for the *DenseNet* and *ResNet* models drop down to 84.38% and 83.66%, respectively, along with dips in all macro-average metrics. We also see a similar degradation in macro-average metrics with the *Efficient* model, but also note how its 'harsh' accuracy is slightly improved to 82.39% from 80.26%. Overall, we view these preliminary metrics in a positive light as the degradations are less than 3% and indicate that the models are most likely training and predicting on new information made available after cropping and equalization.

| Target | Model | Precision (95% CI) | Recall (95% CI) | F1 (95% CI) |
|---|---|---|---|---|
| Normal | DenseNet-121 ImageNet | 94.46% ± 2.68% | 97.50% ± 1.83% | 95.96% ± 2.31% |
| | ResNet-50 ImageNet | 93.81% ± 2.82% | 97.50% ± 1.83% | 95.62% ± 2.40% |
| | Efficient Model | 91.25% ± 3.31% | 96.79% ± 2.06% | 93.93% ± 2.80% |
| Pneumonia | DenseNet-121 ImageNet | 98.32% ± 1.22% | 96.46% ± 1.76% | 97.38% ± 1.52% |
| | ResNet-50 ImageNet | 98.31% ± 1.23% | 95.75% ± 1.92% | 97.01% ± 1.62% |
| | Efficient Model | 97.56% ± 1.47% | 94.34% ± 2.20% | 95.92% ± 1.88% |
| Virus | DenseNet-121 ImageNet | 89.35% ± 3.59% | 82.75% ± 4.39% | 85.92% ± 4.05% |
| | ResNet-50 ImageNet | 86.25% ± 4.01% | 81.69% ± 4.50% | 83.91% ± 4.27% |
| | Efficient Model | 86.26% ± 4.00% | 79.58% ± 4.69% | 82.78% ± 4.39% |
| Bacteria | DenseNet-121 ImageNet | 79.39% ± 6.70% | 74.28% ± 7.24% | 76.75% ± 7.00% |
| | ResNet-50 ImageNet | 73.38% ± 7.32% | 72.86% ± 7.37% | 73.12% ± 7.34% |
| | Efficient Model | 74.40% ± 7.23% | 66.42% ± 7.82% | 70.19% ± 7.58% |
| COVID | DenseNet-121 ImageNet | 100% ± 0% | 91.67% ± 4.51% | 95.65% ± 3.33% |
| | ResNet-50 ImageNet | 100% ± 0% | 91.67% ± 4.51% | 95.65% ± 3.33% |
| | Efficient Model | 100% ± 0% | 91.67% ± 4.51% | 95.65% ± 3.33% |

Table 9. Class-specific metrics of Experiment C model performances

From the target-specific metrics, we find that the most significant changes occurred between differentiating the *Virus* and *Bacteria* labels. Specifically, we find that recall metric, for all models, and the precision metric, for *ResNet* model, degraded for those two classes. Overall, we believe that the degradations we observe in training are most likely due to all models struggling to differentiate these labels and we may benefit from including more samples of specific *Virus* and *Bacteria* caused pathologies, moving forward. What is most interesting is that all models achieve the exact same metrics at detecting the *COVID* target, which hints at the fact that all models are equally capable for this task and dataset.

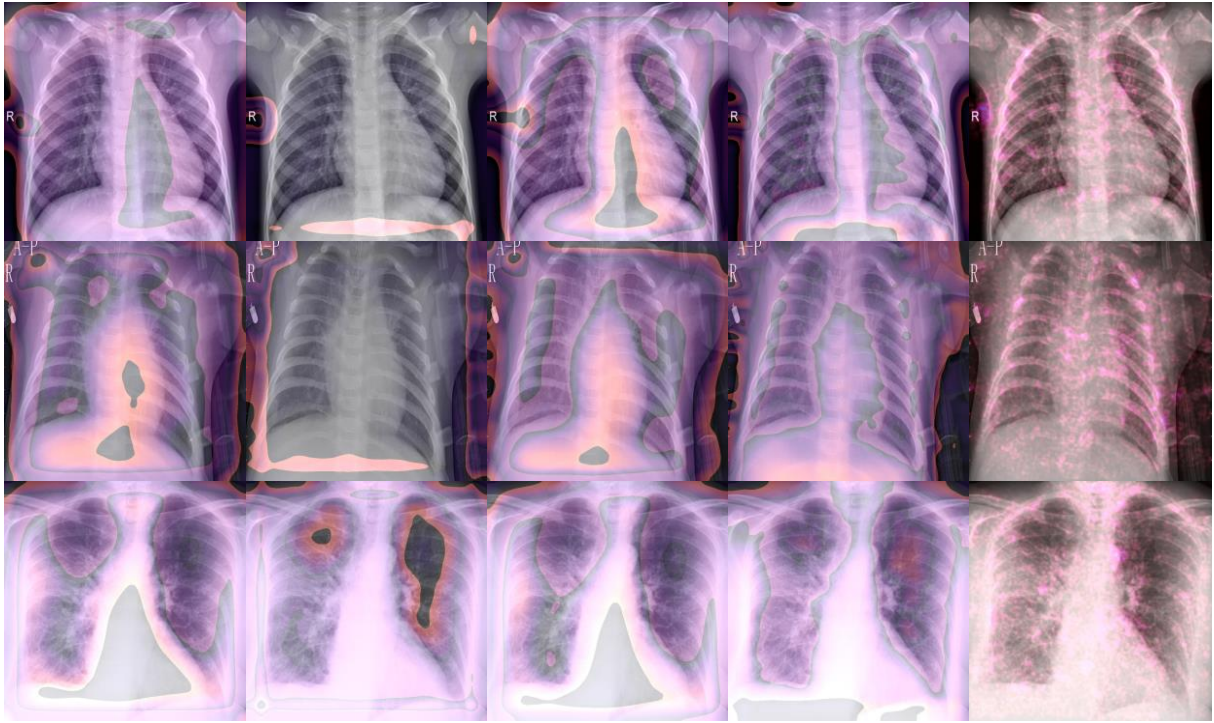

*Figure 11. Rows, Top to Bottom: Normal label, Bacteria label, COVID label.*
*Columns, Left to Right: Unweighted DenseNet, ImageNet DenseNet, Unweighted ResNet, ImageNet ResNet, Efficient model*

## VII. VISUALIZATION

Judging solely from the model metrics, with a few exceptions, all of our models are seemingly capable at detecting *COVID* from chest radiographs and promise very impressive performance. These metrics, however, have to be taken with a grain of a salt as the dataset is not as mature or comprehensive as we would like it to be and we still do not have insight into how or why the models would make a prediction for one target over another. To verify, validate and explore our predictions, we overlay what our models "see" when they are about to make a prediction for an input image, for every image in our test set.

Rajpurkar et al. [2] generated very insightful visualizations, when proposing *CheXNet*, using Class Activation Maps, or CAMs. To generate the CAMs, the images were fed into their fully trained network, the feature maps were extracted from the final convolution layer and the most important features were upscaled to generate a heatmap, displaying the most salient features used to classify an image as a certain pathology. For our *DenseNet* and *ResNet* models, we use a similar method of generating said heatmaps through the use of generated Grad-CAMs, which were proposed to address certain limitations of CAMs by using gradient information flowing into the last convolutional layer of the CNN to understand each neuron for a decision of interest [22].

For our *Efficient* models, however, we use saliency maps to visualize the areas-of-interest of each image in our testing set. Saliency maps, unlike class activation maps, visualize the pixels in an image that contribute the most to predictions by the model through the use of gradient of the predicted outcome from the model with respect to the input values. We use this method for the *Efficient* models because Pasa et al. [4] found that the inclusion of max pooling layers and the small depth of the model cause Grad-CAMs to be of lower resolution and not as useful for diagnostic purposes.

### A. Experiment A

In *Figure 11*, we display examples of visualizations for correctly classified chest radiographs, by all models, for the *Normal*, *Bacteria* and *COVID* targets. From a quick overview, we make some observations that have significant ramifications for how we interpret our results and, subsequently, models.

First, we observe that loading the *DenseNet* and *ResNet* models with the ImageNet weights allow the models, unlike their unweighted counterparts, to pick up on finer details during training and make use of them to make predictions. Unfortunately, this increased ability leads our weighted models to find irrelevant commonalities between targets and use those to make classifications. We see this most clearly with the Grad-CAMs for the *Normal* and *Bacteria* images correctly classified by the *DenseNet* model preloaded with the ImageNet weights: The model made the *Normal* determination using just the "R" orientation maker and something gathered from the abdomen; For the *Bacteria* determination, the model used the tick marks along the edges of the radiograph and some information gleaned from the abdomen, as well. Our *Efficient* model also seems to fall into this trap but manages to, at least, use more than just the markers to make a target determination.

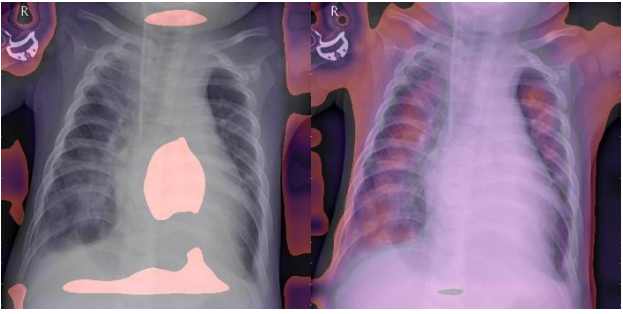

*Figure 12. Comparison of Pneumonia Grad-CAMs for person370_bacteria_1691: DenseNet (left) vs. ResNet (right)*

Second, we find that although both pretrained models tend to pick-up on and be influenced by fine, irrelevant radiograph markers, the weighted *ResNet* model does not make predictions based solely off said irrelevant features and actually manages to identify the desired, relevant features within the lung. We share a typical example of this key difference in *Figure 12*, where we see how the *ResNet* model identifies certain areas-of-interest across the right lung and the top of the left lung, that the *DenseNet* model does not identify, when determining the radiograph is affected by *Pneumonia*.

Finally, and fortunately, we do find confirmation that the *COVID* samples contain features that are distinct enough to allow our models to identify them. In this experiment, we find that these features are typically identified as clusters of varied opacities and veins midway or higher in both lungs.

With this understanding, we find that the ImageNet weighted *ResNet* model is the best performing model in this experiment, as it achieves the highest performance and makes predictions using the relevant radiographic features of our inputs.

### B. Experiment B

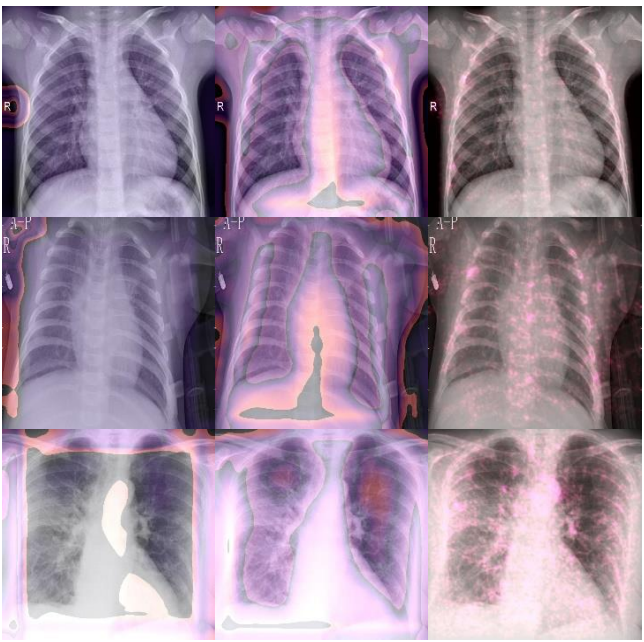

*Figure 13. Rows, Top to Bottom: Normal, Bacteria, COVID. Columns, Left to Right: ImageNet DenseNet, ImageNet ResNet, Efficient model*

In *Figure 13*, we provide visualizations, for the same images we visualized in *Figure 11*, using our *Experiment B* models and we observe some patterns that change how we view the respective performance metrics.

For example, we find that training on a rebalanced dataset was not enough for the *DenseNet* model to train on or predict using features within the lungs; Instead, the model became more prone on using irrelevant features to make determinations, as seen in *Figure 13*. We also find that, for certain samples, the *DenseNet* model seems use the shape of the lungs or chest to make a prediction. We see this most clearly with the example *COVID* Grad-CAM, where the areas with the highest influence are the borders of the lungs. Unfortunately, it seems that in this experiment, the *DenseNet* model is overfitted on the dataset and we cannot expect it to perform well on radiographs outside of this dataset.

Our *ResNet* and *Efficient* models, however, seem to receive some benefit from the rebalanced dataset. We observe better defined Grad-CAMs from our *ResNet* model and cleaner saliency maps from our *Efficient* model. Although, we still see some undesired influence from irrelevant markers, the models seem to benefit from the rebalanced dataset by better identifying and utilizing the desired features.

Again, we determine that the *ResNet* model is the best performing and most reliable model in this experiment, considering how it outperforms the *Efficient* model and the *DenseNet* model is disqualified due to its visualizations.

### C. Experiment C

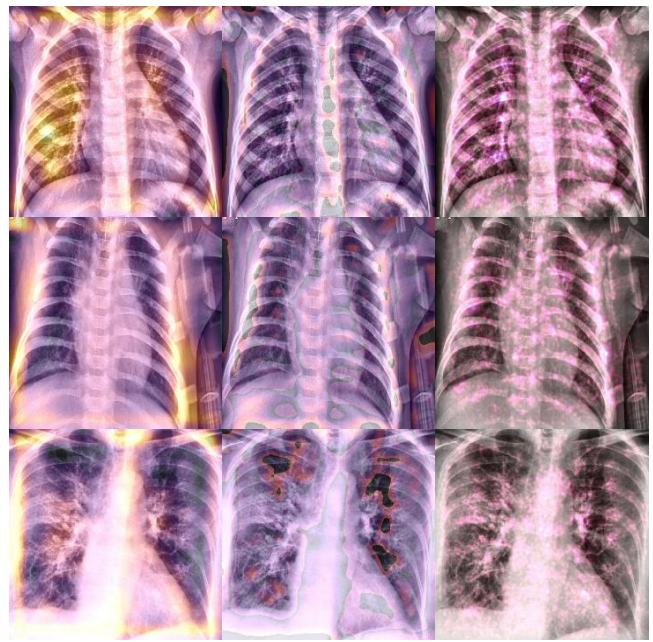

*Figure 14. Rows, Top to Bottom: Normal, Bacteria, COVID. Columns, Left to Right: ImageNet DenseNet, ImageNet ResNet, Efficient model*

Compared to the realized mixed quality of our models and visualizations in *Experiment A* and *Experiment B*, we find that, in *Experiment C*, all of our models saw improvements in identifying relevant features within the lung and using them to make determinations.

We observe the most significant improvement with the *DenseNet* model in *Experiment C*; As seen with the Grad-CAMs for the *Normal* and *COVID* predicted images in *Figure 14*, the *DenseNet* architecture finally was able to identify and use relevant features. This confirms that this architecture requires a large and varied enough dataset to train properly, as it only identified or used relevant features when it trained on our mildly preprocessed dataset, that was heavily augmented during training. Despite this promising improvement, we still observe some legacy behavior and hints of overfitting; Specifically, we observe how the *DenseNet* model also considers the shape of the collarbone, torso and diaphragm when making a prediction.

With regards to the *ResNet* and *Efficient* models, we find that they benefit significantly from the *Experiment C* training methodology and observe these iterations of models considering finer areas-of-interest, than their prior counterparts, when making a prediction. For the *ResNet* model, we see this most clearly between the *Experiment A* and *Experiment B* Grad-CAMs for the *Bacteria* and *COVID* images: In *Experiment A*, the model focused on general areas within the lung, whereas, in *Experiment B*, the model identified finer features to consider and specific areas to exclude. For the *Efficient* model, we find similar improvements in feature identification and localization, and, as a result, find the saliency maps to be clearer, better guided, and more useful for diagnostic purposes.

We believe that the *ResNet* model is again the "best" performing model, as its Grad-CAMs indicate very impressive feature identification and localization capability, it outperforms the *Efficient* model, and the *DenseNet* model still shows signs of being overfit. Compared to the prior determinations, this determination was reached much more narrowly, especially when we observed desired feature identification and usage from all of our visualizations, the gap between the predictive model metrics was significantly narrowed, and all models achieved the same class-specific metrics for detecting the *COVID* target.

## VIII. CONCLUSION

We conclude that, although deep learning methods may be applied to detect *COVID-19* infections from chest radiographs with great promise, current methodologies are insufficient and current proposals that promise no misclassifications should be taken with a grain of salt. We confirm this in *Experiment A,* where we follow contemporary proposed methodologies and achieve very impressive results, by training models on an unprocessed dataset. On a deeper analysis of our results, however, we found that all of our models suffered from using irrelevant markers on the radiographs when making predictions, especially the most proposed models: the deep models preloaded with ImageNet weights.

We did not achieve truly reliable results or model performance until we utilized a relatively simple preprocessing and augmentation training methodology in Experiment C. The gap between our overall model metrics, for example, closed to a range of less than 3%. We also saw significant improvements with relevant feature identification, localization, and utilization from all evaluated models; This is especially true for the ImageNet weighted *DenseNet* model, which failed to identify relevant features in *Experiment A* and *Experiment B*, but now was able to. The lack of discussion regarding any preprocessing in contemporary studies is very concerning and casts doubt on their promised performances, as the models may not be properly trained or simply overfit.

Interestingly, we found that, unlike the *DenseNet* models, the *Efficient* and *ResNet* models were able to identify and utilize relevant features in all of our experiments. Perhaps, this is related to how said models propagate information through the layers: The *ResNet* and *Efficient* models utilize shortcut connections to have subsequent layers consider the residual identity or convolved derivative, respectively, along with the prior layer's gradients during training to maximize information flow; The *DenseNet* model's connections, however, are discriminative and only provide information to subsequent layers with the same feature map size. Simply, this distinction has our *ResNet* and *Efficient* models consider the whole of an input at all stages during training, whereas the *DenseNet* model would tend towards considering the most common, influential features of an input. As such, we find a structural advantage to using architectures similar to the *ResNet* and *Efficient* models, especially when datasets may be relatively small or unbalanced.

From our findings, to develop truly reliable models able to properly detect COVID-19 from radiographs, we make the following suggestions:

**Dataset:** Current contemporary studies suffer and promise potentially misleading performances due to the simple fact that there are not enough publicly available COVID-19 radiograph samples available. Current efforts exist to accomplish this, however, such COVID-19 datasets should be consolidated with existing radiograph datasets, for other pathologies, to develop a holistic, mature dataset. Working on such a dataset will allow researchers and their models to consider how COVID-19 relates to other pathologies, especially when the disease has a high coinfection rate, and allow potential insight into how the virus manifests itself, how distinct or similar it is to other pathologies and how to develop new methods of treating it.

**Preprocessing:** We observed how CLAHE and simple, static cropping could be used to improve the quality of our radiographs and, subsequently, allow our models to better identify and train on the desired relevant features. Such preprocessing should be included in subsequent studies on this subject. Our naive methodology yielded significant improvements, and we believe "smarter" methods may yield even better results; Whether it be from manual or automated cropping to focus on just the lungs, segmentation, or other methods.

**Augmentation:** We have seen how even light augmentation during training yielded significant improvements in feature identification, localization, and utilization in *Experiment C*. As such, augmentation should be used during training as a way to

**Visualization:** As exciting as almost perfect deep learning predictions are, especially with regards to image analysis, an effort should be taken to confirm such predictions were made properly. We found the use of Grad-CAMs and saliency maps very useful in this study, especially when they revealed issues with our results in *Experiments A* and *B*. Such visualization techniques should be utilized in subsequent studies to both validate results and act as a steppingstone for a diagnostic visualization tool.

**Targets:** In related studies, models were trained to perform multi-class classification, where only one target label is predicted for each input. We believe that in the domain of detecting and diagnosing pathologies from chest radiographs, performing multi-class classification would be missing the forest for the trees, especially when COVID-19 has a significant coinfection rate and relationships between this and other pathologies may be lost. For this reason, we recommend that subsequent studies should train their models to perform multi-label classification, where multiple target labels may be predicted for each input; Especially when we have confirmed that models are able to are able to be discriminative, as shown by correctly predicting and differentiating between images that are labeled *Pneumonia, Virus,* and *COVID* and images labeled just *Pneumonia* and *Virus*.

For the sake of transparency and collaboration to take on the challenges presented by the COVID-19 pandemic, we have made all our datasets, scripts, models, results, and visualizations publicly available at: https://github.com/n-albert/COVID-detection-from-radiographs.